\documentstyle[prd,aps,epsf]{revtex}

\newcommand{\beq}{\begin{equation}}
\newcommand{\beqn}{\begin{eqnarray}} 
\newcommand{\eeq}{\end{equation}}
\newcommand{\eeqn}{\end{eqnarray}}
\newcommand{\beqa}{\begin{eqnarray}}
\newcommand{\eeqa}{\end{eqnarray}}
\newcommand{\bea}{\begin{eqnarray}}
\newcommand{\eea}{\end{eqnarray}}

\def\beq{\begin{equation}}
\def\eeq{\end{equation}}
\newcommand{\gsim}{\mbox{\raisebox{-1.ex}{$\stackrel
     {\textstyle>}{\textstyle\sim}$}}}
\newcommand{\lsim}{\mbox{\raisebox{-1.ex}{$\stackrel
     {\textstyle<}{\textstyle \sim}$}}}
\newcommand{\square}{\kern1pt\vbox{\hrule height
1.2pt\hbox{\vrule width 1.2pt\hskip 3pt
   \vbox{\vskip 6pt}\hskip 3pt\vrule width 0.6pt}\hrule
height 0.6pt}\kern1pt}

\begin{document}
\draft \twocolumn[\hsize\textwidth\columnwidth\hsize\csname
@twocolumnfalse\endcsname

\title{Cosmological density perturbations from perturbed couplings} 
\author{Shinji Tsujikawa} 
\address{Institute of Cosmology and Gravitation, 
University of Portsmouth, Mercantile House, 
Portsmouth PO1 2EG, \\
United Kingdom \\[.3em]} 
\date{\today} 
\maketitle
\begin{abstract}
The density perturbations generated when the inflaton decay rate is perturbed 
by a light scalar field $\chi$ are studied.
By explicitly solving the perturbation equations for the 
system of two scalar fields and radiation, we show that even in low 
energy-scale inflation nearly scale-invariant spectra of scalar 
perturbations with an amplitude set by observations are obtained through the 
conversion of $\chi$ fluctuations into adiabatic density perturbations.
We demonstrate that the spectra depend on the average decay rate 
of the inflaton \& on the inflaton fluctuations.
We then apply this new mechanism to string cosmologies \& generalized Einstein 
theories and discuss the conditions under which scale-invariant spectra are 
possible.
\end{abstract}
\vskip 0.1pc 
\vskip 0.1pc
]

In inflationary cosmology the seeds of large-scale structure 
are originated from quantum fluctuations of a light scalar 
field responsible for inflation-- called the {\it inflaton}.  The adiabatic 
perturbations generated from inflaton fluctuations exhibit nearly 
scale-invariant spectra, as required by observations \cite{Spergel:2003cb}.  
The energy scale of inflation is determined by the amplitude of large-scale 
perturbations in the Cosmic Microwave Background (CMB).  
For example, in the simple model of chaotic inflation with potential $V=(1/2)
m_{\phi}^2\phi^2$, the mass of the inflaton $\phi$ is constrained to be 
$m_{\phi} \simeq 10^{-6}m_{\rm pl}$ by the Cosmic Background 
Explorer (COBE) normalization \cite{Linde,LL} 
(Here $m_{\rm pl} \simeq 1.22 \times 
10^{19}\,{\rm GeV}$ is the Planck mass).

On the other hand, if inflation occurs at a much lower scale, the amplitude of 
density perturbations is too low to explain the temperature anisotropy in 
CMB.  In this case we need to consider some other mechanism to generate 
sufficient density perturbations.

Recently a number of authors \cite{Dvali:2003em,Kofman:2003nx}
independently proposed an alternative method to produce scalar metric 
perturbations using the spatial variation of the inflaton 
decay rate, $\Gamma$.  Since the coupling strength is generally a function of 
the vacuum expectation value of scalar fields in string theory, it is 
natural to consider the case where the coupling fluctuates due to scalar 
field perturbations.  When there exists some light field whose mass is 
smaller than the Hubble rate (such as modulus \cite{Kofman:2003nx}), its 
large-scale fluctuations are not exponentially suppressed during inflation.  
This can provide another source of scalar metric perturbations in addition 
to the inflaton.  In fact the authors in \cite{Dvali:2003em} considered the 
decay of the inflaton by treating it as a matter fluid and showed that 
density perturbations can be generated by the perturbed decay rate, $\delta 
\Gamma$.

In this work we shall extend the analysis of Ref.~\cite{Dvali:2003em}
to the more realistic situation where the inflaton is treated as a scalar field,
and completely follow the evolution of metric perturbations during 
inflation and reheating.  This is particularly important when we treat the 
dynamics of reheating consistently, since the fluid approach can lead to 
some loss of information (like parametric resonance) by averaging over 
the scalar field oscillations.  We will analyze the case where 
perturbations of both the inflaton and the decay rate coexist, and estimate the 
amplitude and the spectral index of metric perturbations after inflaton 
decay.  We shall also apply this new mechanism to string-inspired cosmologies 
\& generalized Einstein theories.

Let us consider the chaotic inflationary scenario with 
potential $V=(1/2)m_{\phi}^2\phi^2$.
During reheating, the inflaton $\phi$ decays to 
standard light particles through the interaction, $\lambda_0 \phi \sigma 
\sigma$, whose decay rate, $\Gamma$, is proportional to $\lambda_0^2$ 
\cite{Linde}.  The inflaton can also decay via non-renormalizable 
interactions with superfields, {\it e.g.,} $\phi (q/M)qq$ ($M$ is some 
mass scale), in which case the vacuum expectation value of a scalar 
component of the $q$ superfields, $\langle \chi \rangle$, gives rise to an 
effective coupling, $\lambda=\langle \chi \rangle/M$ \cite{Dvali:2003em}.  
Following Ref.~\cite{Dvali:2003em} we shall analyze the case where the 
effective coupling $\lambda(\chi)$ is dependent on the field $\chi$ as 
\begin{eqnarray}
\lambda(\chi)=\lambda_0 \left(1+\chi/M+ \cdots \right)\,.
\label{lam}
\end{eqnarray}

When the coupling, $(1/2)g^2\phi^2\chi^2$, exists during the reheating 
stage, this can lead to explosive particle production called preheating 
\cite{TB,Kofman:1994rk}.  In order for preheating to occur, we require the 
large resonance parameter, $q_{\rm re} \equiv g^2\phi^2/(4m_{\phi}^2) \gg 
1$, at the end of inflation \cite{Kofman:1994rk}, in which case large-scale 
perturbations in $\chi$ are exponentially suppressed during inflation due to 
the heavy effective mass, $g|\phi|$, relative to the Hubble rate, $H$.
Then the perturbations of the coupling $\lambda(\chi)$ are vanishing small, 
which do not affect the evolution of metric perturbations.  Therefore in this work 
we do not take into account the effect of preheating coming from the 
interaction, $(1/2)g^2\phi^2\chi^2$.  When this interaction is added to the 
Born decay term, $\lambda_0 \phi \chi^2$, this gives rise to the form of 
the effective coupling (\ref{lam}) perturbed by the inflaton $\phi$ instead 
of $\chi$.  However we are interested in the case where the coupling is 
perturbed by a light field $\chi$ other than the inflaton.

Hereafter we will analyze the two-field system of $\phi$
and $\chi$ with radiation.  We assume that the field 
$\chi$ has a small mass, $m_\chi$, relative to the Hubble rate, 
$H \equiv \dot{a}/a$, during inflation (here $a$ is a scale factor).  
Radiation is generated through the decay of the inflaton with the $\chi$-dependent 
coupling (\ref{lam}).  Consider the following perturbed metric for scalar 
perturbations in the longitudinal gauge \cite{KS}: 
\begin{eqnarray}
ds^2 &=& -(1+2\Phi)dt^2 
+a^2(t)(1-2\Psi)\delta_{ij}  dx^i dx^j\,.
\label{pmetric}
\end{eqnarray}
At linear order one has $\Phi=\Psi$ due to the vanishing 
anisotropic stress.  Then the Fourier transformed, linearized 
Einstein equations are \cite{Hwang:2001qk}\footnote{After this work was completed, the authors in 
Ref.~\cite{MR} developed a gauge-invariant formalism for the large-scale curvature
perturbations with spatial and time variations of the inflaton decay rate
using the formalism in Ref.~\cite{Malik}.  
See Ref.~\cite{Mazumdar} for a related work. }  
\begin{eqnarray}
\label{b1}
& &(k^2/a^2)\Phi+3H(\dot{\Phi}+H\Phi) -(4\pi/m_{\rm 
pl}^2)(\dot{\phi}^2+\dot{\chi}^2)\Phi \nonumber \\
& & = -(4\pi/m_{\rm pl}^2) \left[\dot{\phi} 
\delta\dot{\phi}+m_\phi^2\phi \delta\phi+
\dot{\chi}\delta\dot{\chi}+m_\chi^2\chi \delta\chi
+\delta \mu\right], \\
\label{b2}
& &\delta \ddot{\phi}+(3H+\Gamma)\delta \dot{\phi}+\left( 
k^2/a^2+m_{\phi}^2\right)\delta\phi \nonumber \\
& &=4\dot{\phi}\dot{\Phi}-2m_{\phi}^2\phi\Phi
-\Gamma \dot{\phi}(2\Phi +\delta \Gamma/\Gamma), \\
\label{b3}
& &\delta \ddot{\chi}+3H\delta \dot{\chi}+\left( 
k^2/a^2+m_{\chi}^2\right)\delta\chi  = 
4\dot{\chi}\dot{\Phi}-2m_{\chi}^2\chi\Phi, \\
\label{b4}
& &\delta \dot{\mu}+4H\delta \mu =
4\mu \left[\dot{\Phi}-(k/3a)v \right]+ 
\Gamma \dot{\phi} \left[2\delta \dot{\phi}+ 
(\delta \Gamma/\Gamma) \dot{\phi}\right], \\
\label{b5}
& &\dot{v}+(4H+\dot{\mu}/\mu)v= 
(k/a)\left[\Phi+(\delta \mu+3\Gamma \dot{\phi} 
\delta \phi)/4\mu \right],
\end{eqnarray}
where $k$ is a comoving wave number (the $k$ subscript is implicit) 
and $\Gamma$ is the inflaton decay rate that is smaller than the Hubble 
rate during inflation.  $\delta \mu$ is the perturbation of radiation whose energy 
density is given by $\mu$, and $v$ is the velocity of 
the radiation fluid.

The decay rate, $\Gamma$, is proportional to $\lambda^2$,
and $\lambda$ is dependent on the field $\chi$ through the relation (\ref{lam}).  
Therefore the fluctuation of the decay rate can be written as 
\begin{eqnarray}
\frac{\delta \Gamma}{\Gamma}=2\frac{\delta\lambda}{\lambda}
=2\frac{\delta \chi}{\chi+M}\,,
\label{delgamma}
\end{eqnarray}
where we dropped the contributions coming from the higher-order terms
in Eq.~(\ref{lam}). 
When the mass of the field $\chi$ is smaller than $H$, large-scale 
$\chi$ fluctuations are not suppressed during inflation, thereby providing 
source terms for the gravitational potential $\Phi$.

We shall analyze the situation where the background field $\chi$ is not 
dynamically important relative to the inflaton $\phi$ in order to avoid the 
second stage of inflation to occur.  This means that $\dot{\chi} \ll 
\dot{\phi}$ in Eq.~(\ref{b1}), in which case the correlation between 
adiabatic and isocurvature perturbations is small \cite{Gordon:2000hv} due 
to the negligible angular velocity in field space (${\rm tan}\,\theta \equiv 
\dot{\chi}/\dot{\phi} \simeq 0$).  In the absence of the $\delta \Gamma$ 
term the conversion from isocurvature to adiabatic perturbations occurs 
significantly only when $\dot{\chi}$ and $\dot{\phi}$ can be the same order 
during inflation \cite{Tsujikawa:2002qx}.  This corresponds to double 
inflationary scenarios driven by two scalar fields, in which case the 
strength of the correlation is numerically investigated in 
Refs.~\cite{Tsujikawa:2002qx,Langlois:dw} for $\delta \Gamma=0$.

In the presence of the $\delta \Gamma$ fluctuation, the situation 
is different from the one discussed in Ref.~\cite{Gordon:2000hv}.  
First of all, even if $\dot{\chi}$ is much smaller than $\dot{\phi}$, 
the conversion of the $\delta \Gamma$ fluctuation into the gravitational 
potential occurs as seen from Eqs.~(\ref{b1}), (\ref{b2}) and (\ref{b4}).  
As long as the mass of $\chi$ is light and large-scale $\chi$ fluctuations 
are not suppressed during inflation, this naturally leads to the conversion 
to the gravitational potential through the perturbed decay rate, $\delta 
\Gamma$.  The advantage of this scenario is that we do not need to rely on 
two $\chi$-dependent source terms in Eq.~(\ref{b1}) in order to obtain 
sufficient isocurvature perturbations.  In this sense it is also different 
from metric preheating \cite{Bassett} where the growth of the 
$\chi$-dependent terms sources the gravitational potential.

In the absence of the $\delta \Gamma$ term, the amplitude of scalar 
perturbations generated from inflaton fluctuations are estimated as 
$\delta_H=\sqrt{512\pi/75}\,V^{3/2}/(m_{\rm pl}^3|V_\phi|) |_{k=aH}$ 
\cite{LL}.  Substituting the value $\phi \simeq 3m_{\rm pl}$ corresponding 
to the 60 e-foldings before the end of inflation, we get $\delta_H \simeq 
12m_\phi/m_{\rm pl}$.  Making use of the COBE result $\delta_H \simeq 2 
\times 10^{-5}$, the inflaton mass is constrained to be $m_\phi \simeq 1.7 
\times 10^{-6}m_{\rm pl}$.  The spectral index $n$ for the power spectrum, 
${\cal P}_\Phi \equiv \frac{k^3}{2\pi^2}\langle |\Phi_k|^2 \rangle$, is 
estimated as \cite{LL}: 
\begin{eqnarray}
n=1-6\epsilon+2\eta \simeq 1-\frac{4m_\phi^2}{3H^2}\,,
\label{ntilt}
\end{eqnarray}
where $\epsilon \equiv \frac{m_{\rm pl}^2}
{16\pi}\left(\frac{V'}{V}\right)^2 \simeq \frac{m_\phi^2}{3H^2}$ and $\eta 
\equiv \frac{m_{\rm pl}^2}{8\pi}\frac{V''}{V} \simeq \frac{m_\phi^2}{3H^2}$ 
are the slow-roll parameters.  This is a slightly red-tilted spectrum for 
$m_\phi \ll H$.

\begin{figure}
\epsfxsize=3.5in
\epsffile{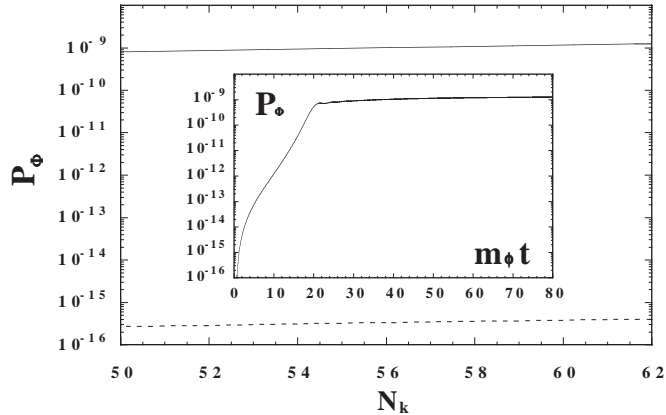} 
\caption{The spectra of the gravitational potential $\Phi$ after inflaton 
decay for $m_{\phi}=10^{-9}m_{\rm pl}$, $m_\chi=0.8m_{\phi}$, 
$M=1.2 \times 10^{-5}m_{\rm pl}$, $\Gamma \simeq 0.1m_\phi$
with initial conditions $\phi_i=3.2m_{\rm pl}$ and $\chi_i=10^{-6}m_{\rm pl}$. 
$N_k$ is the number of e-foldings before the end of inflation. The solid and 
dotted curves correspond to the case with $\delta\Gamma \ne 0$ and 
$\delta\Gamma= 0$, respectively.  We find that the required power spectra, 
${\cal P}_\Phi \sim 10^{-9}$, are obtained by taking into account the $\delta 
\Gamma$ fluctuation.  {\bf Inset}: The evolution of ${\cal P}_\Phi$ for 
$\delta\Gamma \ne 0$ for the mode that crossed the Hubble radius
around $N_k=60$.}
\label{spectra}
\end{figure}

The above basic picture of the generation of density perturbations can be 
substantially modified when the $\delta \Gamma$ fluctuations are involved. 
Let us first estimate the spectral index of the $\delta\chi$ fluctuation in 
Eq.~(\ref{b3}) with mass satisfying $m_\chi~\lsim~H$.  When the field 
$\chi$ is subdominant relative to $\phi$ ($\chi \ll \phi$), we can neglect 
the r.h.s. of Eq.~(\ref{b3}) coming from the gravitational potential.
Then the spectral index for the super-Hubble 
$\delta \chi$ fluctuation is estimated as \cite{Riotto:2002yw} 
\begin{eqnarray}
n_{\delta\chi} =1-\frac{2}{3H^2}(m_\phi^2-m_\chi^2)\,.
\label{ndelchi}
\end{eqnarray}
In the presence of the $\delta \Gamma$ fluctuation, the power 
spectra ${\cal P}_\Phi$ exhibit a spectral tilt which is a mixture of 
Eqs.~(\ref{ntilt}) and (\ref{ndelchi}).  When $m_\chi$ is smaller than 
$m_\phi$, one has slightly red-tilted spectra of $\Phi$ which are close to 
scale-invariant (see Fig.~\ref{spectra}).  The spectra of $\delta\chi$ are 
blue-tilted for $m_\chi>m_\phi$, but this does not necessarily mean that
the gravitational potential is also blue-tilted.
In fact, even when $m_\chi~\gsim~3m_\phi$,
super-Hubble $\delta\chi$ fluctuations begin to be exponentially suppressed
during inflation, in which case the gravitational potential is dominated by
the inflaton fluctuation whose spectrum is given by Eq.~(\ref{ntilt}).
Typically we have red-tilted, nearly scale-invariant spectra of $\Phi$ for 
wide range of the parameter space.

If the inflaton mass is less than $m_\phi \simeq 10^{-6}m_{\rm pl}$, the 
amplitude of the power spectra ${\cal P}_{\Phi}$ is suppressed compared to the 
observed value, ${\cal P}_{\Phi} \simeq 10^{-9}$ (see Fig.~\ref{spectra}).  
Taking into account the $\delta \Gamma$ fluctuation, the amplitude of 
density perturbations can be much higher.  In fact, as shown in 
Fig.~\ref{spectra}, it is possible to get nearly scale-invariant power 
spectra whose amplitude are of order ${\cal P}_{\Phi} \simeq 10^{-9}$.

\begin{figure}
\epsfxsize=3.5in
\epsffile{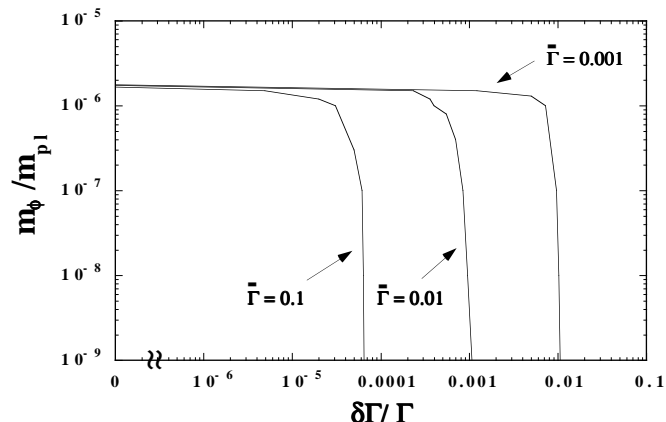} 
\caption{The parameter space where the power spectra, 
${\cal P}_\Phi \simeq 10^{-9}$, are obtained for the mode corresponding to 
$N_k=60$ in the case of $m_\chi=0.1m_{\phi}$.  The condition, $\chi \ll 
M$, is satisfied in all cases of this plot.  Each curve corresponds to the 
case of the average decay rate, $\bar{\Gamma} \equiv \Gamma/m_\phi=
0.1, 0.01$, and $0.001$, respectively.  }
\label{paraplot}
\end{figure}

We also plot in Fig.~\ref{paraplot} the parameter range with respect to  
$m_\phi/m_{\rm pl}$ and $\delta \Gamma/\Gamma$ in which the power-spectra, 
${\cal P}_\Phi \simeq 10^{-9}$, are obtained after inflaton decay.  When 
the $\chi$ mass is light relative to the Hubble rate, one gets the 
perturbations of order $\delta\chi \sim H$ after Hubble radius crossing.  
Therefore the fluctuation of the decay rate is approximately estimated as 
$\delta \Gamma/\Gamma \sim H/M$ for $\chi \ll M$ from Eq.~(\ref{delgamma}).  
We can have a considerable amount of fluctuations in $\delta \Gamma/\Gamma$ 
by varying the mass scale, $M$.  In Fig.~\ref{paraplot} we find that 
inclusion of the perturbed decay rate allows for the wide possibility to 
generate the perturbations of order ${\cal P}_{\Phi} \simeq 10^{-9}$ even 
for $m_\phi \ll 10^{-6}m_{\rm pl}$.  This figure corresponds to the one 
where $\chi$ is much smaller than $\phi$ (and $M$).  We wish to stress that 
$\delta\chi$ fluctuations can be sufficiently transferred to adiabatic 
density perturbations even for ${\rm tan}\,\theta \simeq 0$.  When the 
average decay rate $\Gamma$ gets smaller, we require larger values of 
$\delta \Gamma/\Gamma$ (corresponding to smaller values of $M$) to lead to 
sufficient conversion as confirmed by Eqs.~(\ref{b2}) and (\ref{b4}).  This 
behavior is actually seen in our numerical simulations of 
Fig.~\ref{paraplot}.  In the limit of $\Gamma \to 0$ ($\lambda_0 \to 0$), 
the above conversion does not occur apart from the standard one discussed 
in Ref.~\cite{Gordon:2000hv}.

We can apply the above new mechanism for the more general action 
\begin{eqnarray}
S=\int d^4 x\sqrt{-g} \left[\frac{m_{\rm pl}^2}{16\pi}R-
\frac12 (\nabla \varphi)^2-F(\varphi)(\nabla \chi)^2-U \right],
\label{action}
\end{eqnarray}
where $F(\varphi)$ is a function of $\varphi$ and $U$ is a potential 
of scalar fields including their interactions.
The action (\ref{action}) contains not only various generalized Einstein 
theories \cite{Berkin:nm,Starobinsky:2001xq} 
but also the tree-level Pre-Big-Bang
(PBB) cosmology \cite{PBB,Copeland:1997ug} from the dimensional reduction 
and 4D effective description of ekpyrotic cosmology \cite{Khoury,Cope}.

The effective potential $U$ typically includes an exponential term
in generalized Einstein theories, which can lead to inflationary solutions 
\cite{Berkin:nm,Starobinsky:2001xq}.  Let us consider the soft inflation model 
\cite{Berkin:nm} with $U=U_0 \exp(-\sqrt{2/p}\,\varphi)$ and $F=(1/2) 
e^{\beta \varphi}$ with $\beta$ being a constant (hereafter we set the units 
such that $8\pi/m_{\rm pl}^2 \equiv 1$).  We shall assume that the decay 
rate of the inflaton is perturbed by the field $\chi$ with a negligible 
mass.  The evolution of the background is given as $a \propto t^p 
\propto (-\tau)^{p/(1-p)}$ and $e^{\varphi} \propto \{-(p-1)\tau 
\}^{\sqrt{2p}/(1-p)}$ with $\tau \equiv \int a^{-1}dt\,(<0)$ being a 
conformal time \cite{DiMarco:2002eb}.  Assuming that the background field 
$\chi$ is not dynamically important, the perturbed equation for 
$\delta\chi$ is approximately written as 
\begin{eqnarray}
\label{sigma}
\delta \ddot{\chi}+(3H+\beta\dot{\varphi})\delta\dot{\chi} 
+(k^2/a^2)\delta\chi =0\,.
\end{eqnarray}
Introducing new variables, $b \equiv e^{\beta\varphi/2}a$ and $\delta 
\tilde{\chi} \equiv b \delta \chi$, we get 
\begin{eqnarray}
\label{u}
\delta \tilde{\chi}''+\left(k^2-b'' /b\right) \delta\tilde{\chi}=0\,,
\end{eqnarray}
where a prime denotes the derivative with respect to $\tau$.
When the evolution of $b$ is characterized by 
$b \propto (-\tau)^{\gamma}$, the spectral index of the $\delta\chi$ 
fluctuation is given as $n_{\delta \chi}=4-|1-2\gamma|$ 
\cite{Tsujikawa:2001ad}.  In the present case $\gamma$ is given by 
\begin{eqnarray}
\label{gamma}
\gamma=-\frac{1}{p-1}\left(p+
\beta\sqrt{\frac{p}{2}}\right)\,.
\end{eqnarray}
In the limit of $p \to \infty$, we have $\gamma \to -1$, thereby 
yielding the scale-invariant spectra, $n_{\delta \chi}=1$.  In the presence 
of the noncanonical kinetic term ($\beta \ne 0$), we have $n_{\delta 
\chi}=1$ for $\beta=-\sqrt{2/p}$ or $\beta=-(3p-2)\sqrt{2/p}$ (corresponding 
to $\gamma=-1$ or $\gamma=2$).  If the decay rate of the inflaton is 
perturbed by the field $\chi$, it is possible to generate scale-invariant 
spectra of density perturbations in a similar way as discussed previously.

In PBB and ekpyrotic cosmologies, one has the contracting phase of the 
universe in the Einstein frame before the bounce \cite{Tsujikawa:2002qc}.  
Let us consider the ekpyrotic scenario whose potential is given by $U=-U_0 
\exp(-\sqrt{2/p}\,\varphi)$ with $U_0<0$ and $0<p<1$.  In this 
case the contracting phase is characterized by the scale factor, $a \propto 
(-t)^p \propto (-\tau)^{p/(1-p)}$, with $t<0$.  Since the brane modulus, 
$\varphi$, evolves as $e^{\varphi} \propto \{(1-p)(-\tau)\}^{\sqrt{2p}/(1-p)}$, 
we get the same $\gamma$ given by Eq.~(\ref{gamma}) 
for the fluctuation $\delta \chi$.  In the case of minimal coupling 
($\beta=0$), one has $n_{\delta \chi}= (3-p)/(1-p) \simeq 3$ for $p \simeq 
0$.  This is the same blue spectral index as the curvature perturbation, 
${\cal R}$, obtained in the system of the brane-modulus only 
\cite{Lyth:2001pf}.  We can have $n_{\delta \chi}=1$ for $p=2/3$ 
\cite{Finelli:2001sr}, but the system is found to be unstable in this case 
\cite{Tsujikawa:2002qc}.  When $\beta=-\sqrt{2/p}$ or 
$\beta=(2-3p)\sqrt{2/p}$, metric perturbations can be scale-invariant 
provided that $\delta \chi$ fluctuations are sufficiently converted to 
$\Phi$ through the perturbed decay rate.

The PBB scenario corresponds to the vanishing potential with $p=1/3$.
It is convenient to use the relation $\phi=-\sqrt{2}\varphi$ between the 
dilaton $\phi$ in the PBB case and the field $\varphi$ 
in the ekpyrotic case \cite{Tsujikawa:2002qc}.  This is equivalent to  
replacing $\beta$ in Eq.~(\ref{action}) for new beta $\tilde{\beta}$, as 
$\beta=-\sqrt{2}\tilde{\beta}$.  Then the spectral index for $\delta\chi$ 
is given by $n_{\delta\chi}=4-\sqrt{3}|\tilde{\beta}|$, which allows 
for scale-invariant spectra for $|\tilde{\beta}|=\sqrt{3}$.  Since 
$|\tilde{\beta}|=2$ for the standard axionic coupling, we have 
$n_{\delta\chi} \simeq 0.54$ in this case \cite{Copeland:1997ug}.  If we 
take into account the contribution of the modulus kinetic term, $-n(\nabla 
\alpha)^2$, in the action (\ref{action}) (here $n$ is the number of 
compactified dimensions), it is possible to obtain scale-invariant spectra 
even for $\tilde{\beta}=2$ \cite{Copeland:1997ug}.  When the modulus 
$\alpha$ evolves as $e^{\alpha} \propto (-\tau)^s$, we get the spectral 
index $n_{\delta\chi}=4-|\tilde{\beta}|\sqrt{3-2ns^2}$.  The 
scale-invariant spectra follow for $|\tilde{\beta}|\sqrt{3-2ns^2}=3$.

Of course one must make a numerical analysis in order to check the 
efficiency of the conversion to adiabatic density perturbations.  In 
string-inspired cosmologies with a contracting phase, noncanonical kinetic 
terms ($\beta \ne 0$) are essentially important to yield scale-invariant 
spectra in $\Phi$ through the perturbed decay rate coming from $\delta 
\chi$.  The correlated adiabatic and isocurvature perturbations were 
discussed in Ref.~\cite{DiMarco:2002eb} for the same action (\ref{action}) 
when $\delta\Gamma=0$, implying the possibility to generate scale-invariant 
spectra in entropy field perturbations for $0<p<1$.  In addition to this, 
the perturbed decay coupling provides us a new source for isocurvature 
perturbations, whose conversion to adiabatic perturbations is efficient 
even when the background field $\chi$ is dynamically negligible.

In PBB and ekpyrotic cosmologies, we have radiation just after the bounce 
through the decay of scalar fields.  The spectra of perturbations 
can be affected by the details of the background evolution around the 
bounce as was recently analyzed in Ref.~\cite{Cartier:2003jz}.
It is certainly of interest to investigate the final spectra of $\Phi$ 
for $\delta\Gamma \ne 0$ by passing through the bounce numerically.
Not only in the case of standard inflation, this can also provide us an 
exciting possibility to generate nearly scale-invariant density 
perturbations with the amplitude required by observations in the context of 
string cosmologies.

\underline{\em Acknowledgements} --
The author is grateful to Bruce Bassett and David Parkinson for useful 
discussions and comments.  He also thanks Anupam Mazumdar 
for drawing attention to the references \cite{Enqvist}.  This work is 
supported from JSPS (No.\ 04942).  



\begin{thebibliography}{99}

\bibitem{Spergel:2003cb}
D.~N.~Spergel {\it et al.},
astro-ph/0302209.

\bibitem{Linde} 
A.  Linde, {\em Particle Physics and Inflationary Cosmology},
Harwood, Chur (1990).

\bibitem{LL}
A.~R.~Liddle and D.~H. ~Lyth, {\em Cosmological inflation and
large-scale structure}, Cambridge University Press (2000).

\bibitem{Dvali:2003em}
G.~Dvali, A.~Gruzinov and M.~Zaldarriaga,
astro-ph/0303591.

\bibitem{Kofman:2003nx}
L.~Kofman,
astro-ph/0303614.

\bibitem{TB}
J.~H.~Traschen and R.~H.~Brandenberger,
Phys.\ Rev.\ D {\bf 42}, 2491 (1990);
Y.~Shtanov, J.~H.~Traschen and R.~H.~Brandenberger, 
Phys.\ Rev.\ D {\bf 51}, 5438 (1995).

\bibitem{Kofman:1994rk}
L.~Kofman, A.~D.~Linde and A.~A.~Starobinsky,
Phys.\ Rev.\ Lett.\  {\bf 73}, 3195 (1994);
Phys.\ Rev.\ D {\bf 56}, 3258 (1997).

\bibitem{sup}
K.~Jedamzik and G.~Sigl, Phys. \ Rev. \ D {\bf 61}, 
023519 (2000);
P.~Ivanov, Phys. \ Rev. \ D {\bf 61}, 023505 (2000).

\bibitem{KS}
H. Kodama and M. Sasaki, 
Prog. Theor. Phys. Suppl. {\bf 78}, 1 (1984);
V.~F. Mukhanov, H.~A.~Feldman and 
R.~H.~Brandenberger,
Phys. Rep. {\bf 215}, 293 (1992).

\bibitem{Hwang:2001qk}
J.~Hwang and H.~Noh, 
Phys.\ Rev.\ D {\bf 65}, 023512 (2002).

\bibitem{MR}
S.~Matarrese and A.~Riotto,
astro-ph/0306416.

\bibitem{Malik}
K.~A.~Malik, D.~Wands and C.~Ungarelli,
Phys.\ Rev.\ D {\bf 67}, 063516 (2003).

\bibitem{Mazumdar}
A.~Mazumdar and M.~Postma,
astro-ph/0306509.

\bibitem{Gordon:2000hv}
C.~Gordon {\it et al.}, 
Phys.\ Rev.\ D {\bf 63}, 023506 (2001).

\bibitem{Tsujikawa:2002qx}
S.~Tsujikawa, D.~Parkinson and B.~A.~Bassett,
Phys.\ Rev.\ D {\bf 67}, 083516 (2003).

\bibitem{Langlois:dw}
D.~Langlois,
Phys.\ Rev.\ D {\bf 59}, 123512 (1999).

\bibitem{Bassett}
B.~A.~Bassett, D.~I.~Kaiser and R.~Maartens,
Phys.\ Lett.\ B {\bf 455}, 84 (1999);
B.~A.~Bassett {\it et al.},  
Nucl.\ Phys.\ B {\bf 561}, 188 (1999).
B.~A.~Bassett and F.~Viniegra,
Phys.\ Rev.\ D {\bf 62}, 043507 (2000);
F.~Finelli and R.~H.~Brandenberger,
Phys.\ Rev.\ D {\bf 62}, 083502 (2000);
B.~A.~Bassett {\it et al.}, 
Nucl.\ Phys.\ B {\bf 622}, 393 (2002);
S.~Tsujikawa and B.~A.~Bassett,
Phys.\ Lett.\ B {\bf 536}, 9 (2002).

\bibitem{Riotto:2002yw}
A.~Riotto,
hep-ph/0210162.

\bibitem{Berkin:nm}
A.~Berkin and K.~Maeda, 
Phys.\ Rev.\ D {\bf 44}, 1691 (1991).

\bibitem{Starobinsky:2001xq}
A.~A.~Starobinsky, S.~Tsujikawa and J.~Yokoyama,
Nucl.\ Phys.\ B {\bf 610}, 383 (2001).

\bibitem{PBB}
G.~Veneziano,
Phys.\ Lett.\ B {\bf 265}, 287 (1991); 
M.~Gasperini and G.~Veneziano,
Astropart.\ Phys.\  {\bf 1} (1993) 317.

\bibitem{Copeland:1997ug}
E.~J.~Copeland, R.~Easther and D.~Wands,
Phys.\ Rev.\ D {\bf 56}, 874 (1997).

\bibitem{Khoury}
J.~Khoury {\it et al.},
Phys.\ Rev.\ D {\bf 64}, 123522 (2001).
\bibitem{Cope}
E.~J.~Copeland, J.~Gray and A.~Lukas,
Phys.\ Rev.\ D {\bf 64}, 126003 (2001).

\bibitem{DiMarco:2002eb}
F.~Di Marco, F.~Finelli and R.~Brandenberger,
Phys.\ Rev.\ D {\bf 67}, 063512 (2003).

\bibitem{Tsujikawa:2001ad}
S.~Tsujikawa,
Phys.\ Lett.\ B {\bf 526}, 179 (2002).

\bibitem{Tsujikawa:2002qc}
S.~Tsujikawa, R.~Brandenberger and F.~Finelli,
Phys.\ Rev.\ D {\bf 66}, 083513 (2002).

\bibitem{Lyth:2001pf}
D.~H.~Lyth,
Phys.\ Lett.\ B {\bf 524}, 1 (2002).

\bibitem{Finelli:2001sr}
F.~Finelli and R.~Brandenberger,
Phys.\ Rev.\ D {\bf 65}, 103522 (2002).

\bibitem{Cartier:2003jz}
C.~Cartier, R.~Durrer and E.~J.~Copeland,
Phys.\ Rev.\ D {\bf 67}, 103517 (2003).

\bibitem{Enqvist}
K.~Enqvist, A.~Mazumdar and M.~Postma,
Phys.\ Rev.\ D {\bf 67}, 121303 (2003);
A.~Mazumdar, hep-ph/0306026.  

\end{thebibliography}
\end{document}